\documentclass[a4paper,10pt]{article}
\usepackage[utf8]{inputenc}

\usepackage[top=2.0cm, bottom=2.0cm, left=2.5cm, right=2.5cm]{geometry}
\usepackage{multicol}

\usepackage{amsmath,amssymb}
\usepackage{graphicx}
\usepackage{textcomp}

\title{Annular cracks in thin films of nanoparticle suspensions drying on a fiber\footnote{Accepted in EPL.}}
\author{Fran\c{c}ois Boulogne\textsuperscript{1,2}, Ludovic Pauchard\textsuperscript{1}, Fr\'ed\'erique Giorgiutti-Dauphin\'e\textsuperscript{1}}

\begin{document}

\maketitle
\emph{\textsuperscript{1} UPMC Univ Paris 06, Univ Paris-Sud, CNRS, F-91405. Lab FAST, Bat 502, Campus Univ, Orsay, F-91405, France.\\
\textsuperscript{2} boulogne@fast.u-psud.fr 
}

\begin{abstract}
We report an experimental study of the crack pattern formed during the drying of a colloidal suspension.
A horizontal fiber, which provides a one dimensional, boundary-free substrate, is coated by a film of micronic thickness.
The geometry imposes a remarkable annular crack pattern and allowing precise measurements of the crack spacing over a short range
of film thickness (between 2 and 10 $\mu$m) which varies linearly with the film height.
We compare our experimental data with a model proposed by Kitsunezaki which suggests that the variation of the crack spacing with the film thickness depends on the ratio between a critical stress at cracking and a critical stress for slipping on the substrate. 
By measuring the friction force of the colloidal gels on a hydrophobic surface through a cantilever technique, we can deduce the critical crack stress for these colloidal gels simply by measuring the crack spacing of the pattern.
\end{abstract}

\begin{multicols}{2}
Crack patterns arising from desiccation of a suspension can exhibit some various and aesthetic morphologies.
These can be observed in a variety of systems in a large range of scales from dried soils to paintings where cracks are undesirable.
During the last decade, the general problem of the drying of colloidal dispersions and of the resulting crack patterns has attracted much theoretical and experimental interest \cite{Groisman1994,Allain1995,Routh2001,Lee2004,Tirumkudulu2005,Gauthier2007,Smith2011}.
Although different mechanisms have been identified and known to be responsible for crack formations, there is still much debate and questioning about fracture patterns and their dynamics.
Consensus has been reached concerning the basic mechanisms responsible for cracking: the evaporation of the solvent tends to contract the solid particles through capillary pressure whereas the adhesion of the gel with the substrate resists horizontal shrinkage.
When a critical stress is reached, cracks invade the material.
The resulting crack pattern exhibits a spatial periodicity that scales with the film thickness and strongly depends on the type of experiments.
Two drying geometries are usually carried out resulting in directional cracks propagation (sessile drop \cite{Pauchard1999,Smith2011} or thin cell \cite{Allain1995,Dufresne2003,Gauthier2007}) or isotropic patterns\cite{Lee2004, Bohn2005}.
However the crack periodicity is more difficult to quantify due to the thickness gradients (in sessile drops) or complex patterns which arise in isotropic drying (thin films in Petri dishes far from the border) due to a hierarchical crack formation.
Groisman \textit{et al.} \cite{Groisman1994} have shown that the final number of cracks obtained in a desiccation experiment decreases if the friction between the material and the substrate is lowered.
Numerous models have been proposed to describe the features of crack morphology of nanoparticles.
In particular a linear relation between the scale of the final crack pattern and the thickness of the layer have been proposed by Groisman \textit{et al.} \cite{Groisman1994} and Allain \textit{et al.} \cite{Allain1995}.
For confined geometries, Komatsu \textit{et al.} \cite{Sasa1997} modelised the gel with a network of springs connected to each other and to the substrate, assuming a perfect adhesion on the later.
The condition for the material to break is based on a balance between the elastic energy stored in the material and the energy required to create new surfaces (Griffith criterion \cite{Griffith1920}).
Kitsunezaki \cite{Kitsunezaki1999} proposed a model which supposes friction with the substrate and the accounting of a critical value for the crack stress of the material to recover Groisman results in systems with a free surface.

In this letter, we study the crack pattern resulting from the drying of a colloidal dispersion in a well-controlled experiment which exhibits no boundary effects and presents a one dimensional geometry which makes the problem amenable to theoretical analysis.
This is achieved by considering the drying of a film of silica dispersion coating on a horizontal fiber.
In the case of a flat film thickness, the drying process causes the appearance of annular cracks on the fiber with a well-defined spacing. 
Contrary to other known open systems, this geometry exhibits only one generation of cracks, with a flat thickness. Thus precise measurements of both the crack spacing and the film thickness can be
carried out. We have performed experiments with three systems with different mechanical properties.
This crack spacing is found to be linear with the film thickness, with a slope dependent on the mechanical properties of the gels.
We have compared our experimental data with the Kitsunezaki model predictions.
The linear dependence between the crack spacing and the film thickness can be expressed as a ratio between two critical stresses, i.e the critical stress at cracking and the critical sliding stress on the substrate. In parallel, we proceeded to measurements of the friction force with a hydrophobic substrate for these colloidal gels with a cantilever technique.
These complementary experimental data provide an estimation of the critical stress at cracking in colloidal gels.
These experiments offer an excellent opportunity to investigate the role of both gel cohesion and gel friction properties on the crack mechanism.

\section{Experimental}
\label{experimental}
\begin{figure}
\centering
\resizebox{0.5\columnwidth}{!}{
  \includegraphics{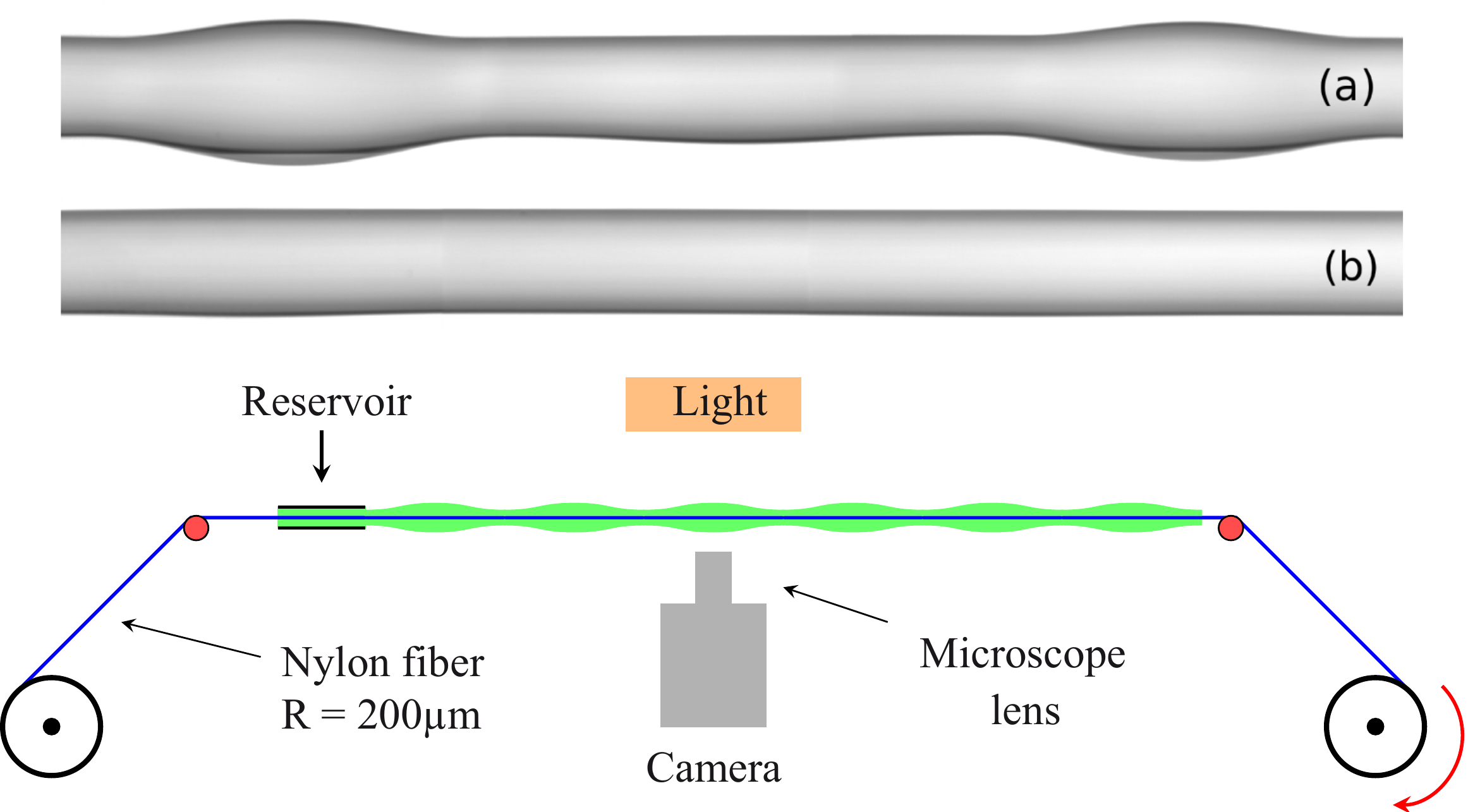}
}
\caption{Top: Sketch of a fiber passing through a colloidal dispersion bath. Observations use a light transmission technique. Bottom: Photographs of fiber coatings before cracking. 
For $h_0 = 33$ $\mu$m $> h_0^c$ (a), the Rayleigh-Plateau instability develops whereas for $h_0 = 18$ $\mu$m $< h_0^c$, the drying process occurs first.
}
\label{fig:exp_setup}
\end{figure}

A sketch of the experimental set-up is shown in figure \ref{fig:exp_setup}.
Experiments are performed by drawing a Nylon fiber (radius $R=200  \mu$m) out of a fluid reservoir at a constant velocity.
The reservoir (a PET tube of length $7$ cm with an inner diameter of $1.6$ mm) contains an aqueous colloidal dispersion of silica particles, Ludox HS-40 (poly-disperse spheres with a radius $7\pm2$ nm and a volume fraction $\phi_0 = 22$\%, purchased from Sigma-Aldrich).
We measured the viscosity $\eta \sim 15$ mPa.s and the surface tension $\gamma\simeq 66$ mN/m.
During drying, particles concentrate and a gel phase is formed at an ambient relative humidity $R_{h} = 50\%$.
The change in the mechanical properties of the gel is achieved by adding a non-volatile compound to the solvent \cite{Boulogne2012a}.
We choose glycerol for the high miscibility in water and for its low vapor pressure ($2.2 \times 10^{-4}$ mbar at $25^\circ$C)\cite{Daubert1989}.
To prepare these solutions, a fraction of Ludox solvent is evaporated and replaced by the corresponding mass of glycerol such as the silica mass fraction remains equal to $40$\%. The glycerol concentration is calculated from the ratio of the mass of glycerol and the mass of the final solution.
Three solutions are studied in this work: $0$, $1.5$ and $3.1$\% (wt) denoted by $\textrm{sol}_{1}$, $\textrm{sol}_{2}$ and $\textrm{sol}_{3}$.

After pulling the wire, the motor is stopped and observations are made with a microscope lens ($15\times$) adapted to a camera (AVT's Marlin) providing digitized images with a resolution about $0.5$ $\mu$m/pixel.
The camera is mounted on a micrometric assembly allowing a precise alignment of the field of view with the film.

\end{multicols}

\begin{figure}
\centering
\resizebox{0.5\columnwidth}{!}{
  \includegraphics{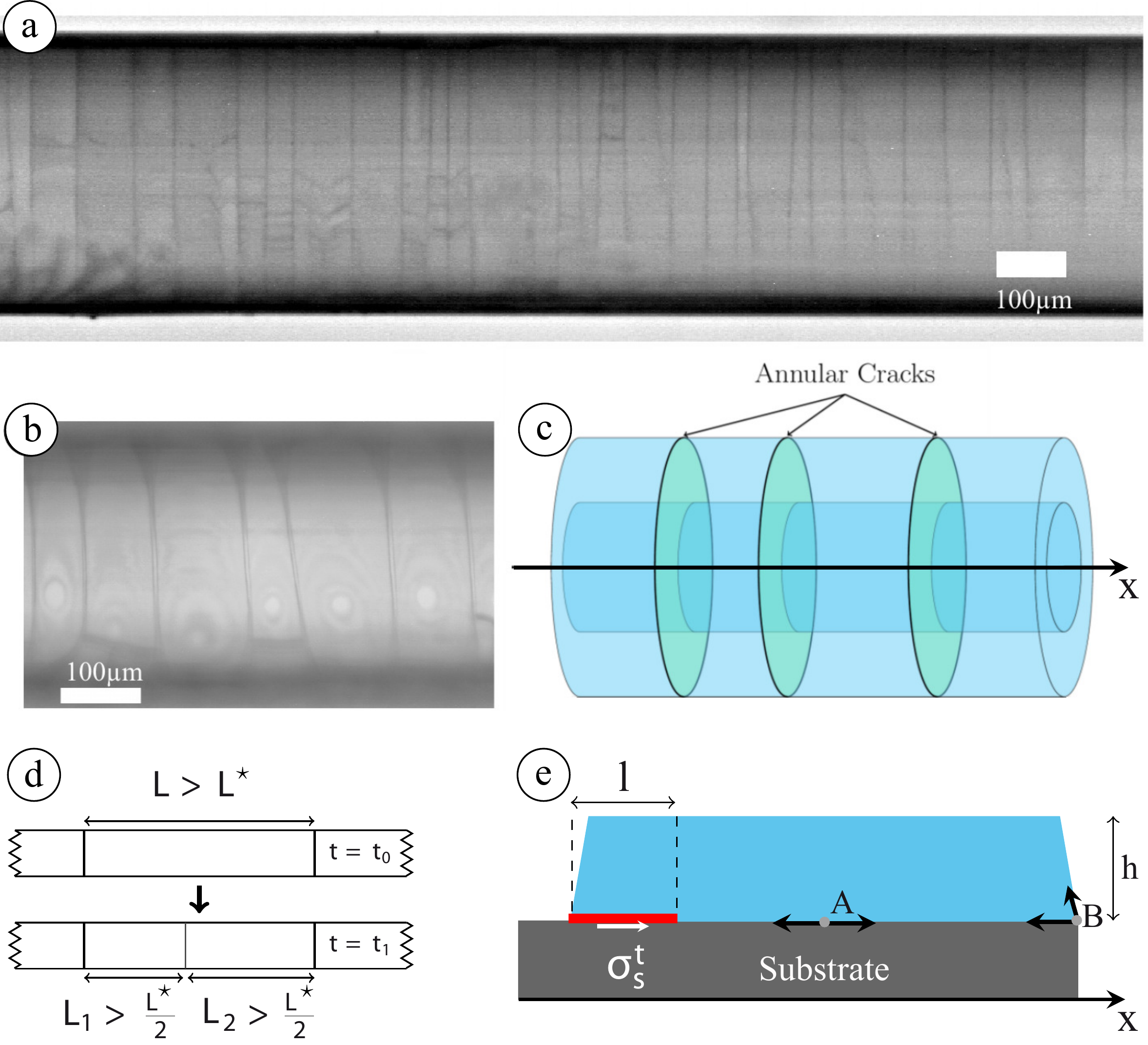}
}
\caption
{(a) Global view of annular crack pattern developed on the coated fiber.
(b) Zoom.
(c) Sketch of the annular cracks.
(d) Schematic representation of cracks generated on the fiber coating: at  $t = t_0$, the crack spacing between the two existing cracks is $L$; at $t = t_1>t_0$, a new crack propagates at a random position between the two existing cracks accordingly with the parking mechanism.
(e) Cross section of a fragment; a sliding friction has a constant shear stress $\sigma^t_s$ resulting in a crack length $l$. }

\label{fig:montage}
\end{figure}

\begin{multicols}{2}

\section{Results}
\label{results}

The initial annular thread coating the fiber can undergo the capillary driven Rayleigh-Plateau instability and exhibits a
 regular pattern of drops separated by a thin liquid film \cite{Plateau1873}.
During the drying process, the material consolidation proceeds and the gel elasticity competes with the instability \cite{Mora2010}.
The evaporation rate $V_e$ is governed by the diffusion-limited transport of water vapor:
$V_e = \frac{D}{R} \frac{n_{sat}}{n_l}(1-R_{h})$ where $D = 2.6\times 10^{-5} \textrm{m}^2\textrm{/s}$ is the diffusion coefficient of water into air, $n_{sat} = 0.91\textrm{mol/m}^{3}$ is the saturation vapor concentration, $n_l = 55000 \textrm{mol/m}^{3}$ is the liquid water concentration.
The radius $R$ scales the vapor density gradient \cite{Dufresne2003,Cazabat2010}.
Since the evaporation causes the film thickness to decrease, the particle volume fraction increases as: $\phi(t) = \frac{\phi_0}{1-V_et/h_0}$.
For our silica dispersion, both viscoelasticity characterization \cite{DiGiuseppe2012} and Small Angle X-ray Scattering study \cite{Li2012a} reflect a jamming of the particles for $\phi^c \simeq 0.35$.
As a result we can estimate the consolidation time $\tau_{cons}$ as $\phi(\tau_{cons})=\phi^c$.
In addition, the instability occurs at a timescale  $\tau_{ins} = \frac{\eta R^4}{\gamma h_0^3}$ where $h_0$ is the film thickness after coating.
The increase of viscosity due to the drying can be expressed by the Krieger-Dougherty law: $\eta(t) = \eta_0 \large( 1 - \phi/\phi_m\large)^{-2.5\phi_m}$ where $\phi_m \simeq 0.60$ \cite{Krieger1959}.
As a result, $\tau_{cons}\sim \tau_{ins}$ yields a critical thickness $h_0^c \simeq 20$ $\mu$m below which the instability is inhibited by the drying process.

The final critical film thickness can be evaluated by $h^c = h_0^c \phi_0 / \phi_m = 7$ $\mu$m which is in a good agreement with experimental observations: this regime is fulfilled for final thicknesses varying in $[2.5,8]$ $\mu$m for which we obtained weak thickness variations ($\pm1.5$ $\mu$m over several centimeters for the highest studied thickness).

The aspect ratio of the fiber, $R/{\cal L} \ll 1$ , with ${\cal L}$ the film length, imposes that the axial stress is larger than the azimuthal stress, thereby leading preferentially to annular cracks formation (figure \ref{fig:montage}-a,b,c).
As illustrated in figure \ref{fig:montage}-d, a first annular fracture actually appears on the fiber and then a second at a distance from the first one higher than the typical size $L^\star$, of the zone relieved of stress around a crack.
The third crack will appear preferentially in between the two first ones but defects in the material will induce a dispersion in the positions of the fractures.
Thus, the distribution of crack spacing is about conform to the basis expected distribution of a random car parking problem \cite{Renyi1958} (figure \ref{fig:montage}-d).

\begin{figure}
\centering
\resizebox{0.5\columnwidth}{!}{
  \includegraphics{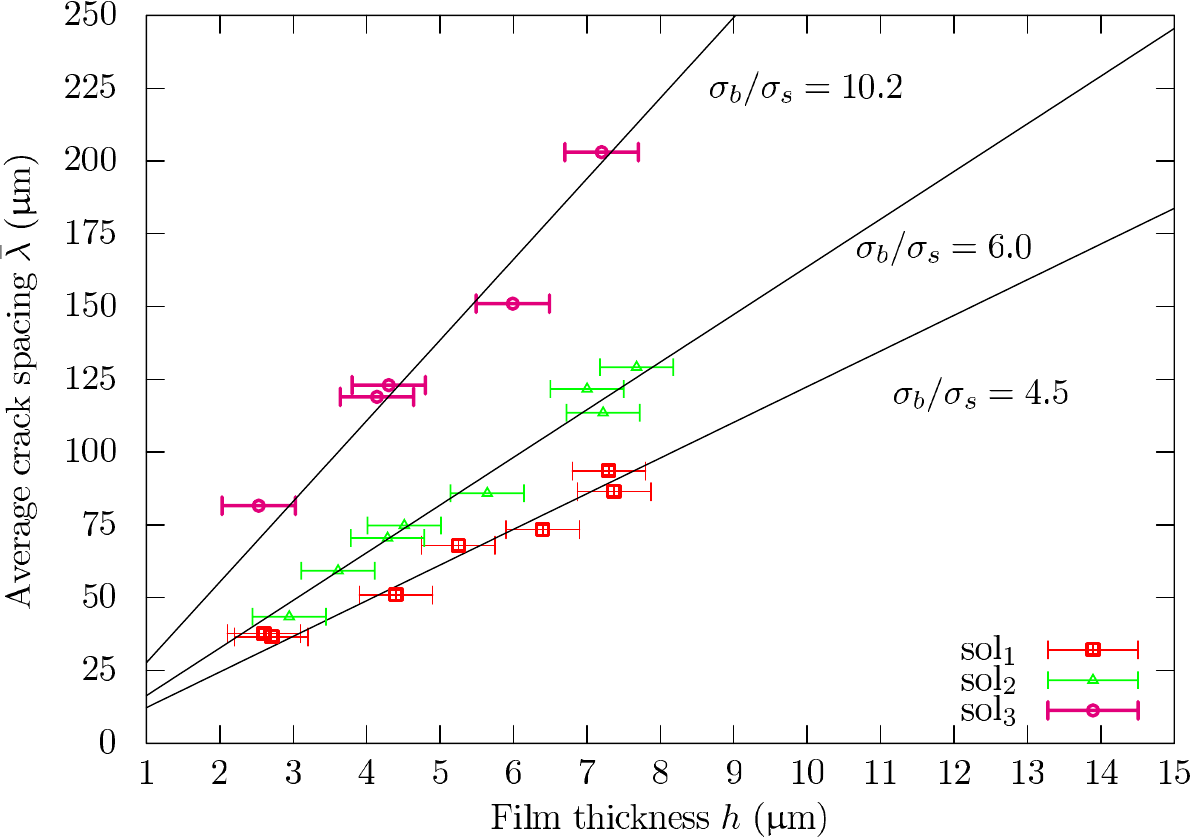}
}
\caption{Average crack spacing $\bar{\lambda}$ as a function of dry film thickness $h$ for three systems. 
Black lines represent fits using equation (\ref{eq:lambda_bar}); the corresponding ratio $\frac{\sigma_b}{\sigma_s}$ is reported for each of them.}
\label{fig:wavelength}
\end{figure}

A typical crack pattern after a complete drying, is presented in figure \ref{fig:montage}-a.
One can note that only one generation of cracks is observed.
We measured approximately $50$ crack spacings per film thickness values ($\pm0.5$ $\mu$m).
The thickness $h$ of the final material is deduced from the measurements of the diameter of each gel slab.
The mean crack spacing $\bar{\lambda}$ as a function of the dry film thickness, $h$, is reported in figure \ref{fig:wavelength} for the three systems prepared with different glycerol concentrations. The choice of glycerol has been motivated by the results of previous works \cite{Boulogne2012a} as it has been evidenced that glycerol is a good candidate to change the mechanical properties of these systems, keeping a constant surface tension.
As expected for open geometries, it appears that the crack spacing increases linearly with $h$.
The slope of the linear fit depends strongly on the glycerol concentration: an addition of $3.1\%$ in glycerol leads to an increase in the slope of $127\%$.
In an attempt to link this slope to physical parameters of the material, we consider a one-dimensional model proposed by Kitsunezaki which includes the general possibility for the gel to slip on the substrate.

\end{multicols}

\begin{multicols}{2}
By modeling the gel as an isotropic and linear elastic material, Kitsunezaki studied the crack spacing resulting from the drying process \cite{Kitsunezaki1999}.
The author considers an isolated fragment submitted to shrinkage, higher near the interface with air than with the substrate (which suggests a cross section of a trapezoidal shape, see figure \ref{fig:montage}-e). Since the maximum of the tensile stress occurs at the border of the fragment (point B), the possibility of a slipping mechanism at this location is considered 
rather than a fixed boundary condition (point A for example). 
The model suggests then that the fragment starts to slip with the substrate starting from point B,  before the fragment is divided with a crack perpendicular to the interfaces. 
Thus, the cracking mechanism results in the competition between two characteristic stresses: a macroscopic breaking stress above which a crack propagates in the material and the sliding stress.
The friction force $F_s$ of the gel on the fiber is assumed to be constant and can be related to the stress $\sigma_s$ in the material by a shear lag model $F_s  \equiv 2\pi R l \sigma^t_s = \sigma_s 2 \pi R h$; therefore $\sigma^t_s = \frac{h}{l}\sigma_{s}$ where $l$ is the contact sliding length as shown in figure \ref{fig:montage}-e\cite{Suo2001}.
Minimizing the elastic energy, Kitsunezaki derived the largest crack pattern length: $L_{max}=  \frac{\sigma_b}{\sigma_s} h_0$.
The relation between the initial thickness $h_0$ and the measured thickness $h$ of the dry film is assumed to be $h = h_0 \phi_0 / \phi_m$.
It results $h_0 \simeq 2.7 h$.
Therefore, the largest predicted crack spacing $\lambda_{max}$ is given by $\lambda_{max} \sim 2.7 \frac{\sigma_b}{\sigma_s} h$.
Assuming that the mean crack spacing is $\bar{\lambda}\simeq \lambda_{max}$, this equation becomes:

\begin{equation}
\bar{\lambda} \sim 2.7 \frac{\sigma_b}{\sigma_s} h
\label{eq:lambda_bar}
\end{equation}

The experimental data are well fitted by relation (\ref{eq:lambda_bar}) and the ratio $\frac{\sigma_{b}}{\sigma_{s}}$ is much larger than one, assuring the model consistency and confirming the sliding motion on the substrate before cracking.

\end{multicols}
\begin{figure}
\centering
\resizebox{0.5\columnwidth}{!}{
  \includegraphics{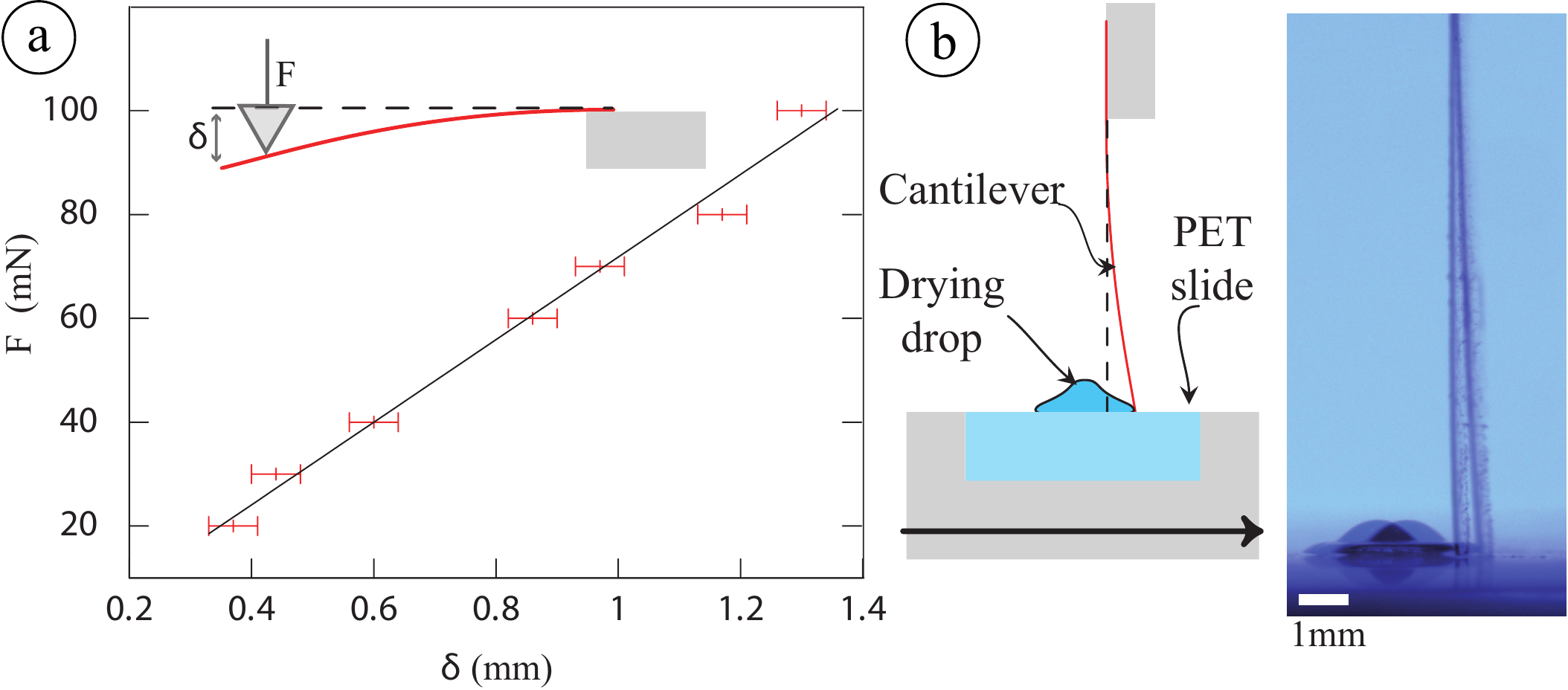}
}
\caption{(a) Calibration method of the cantilever: a force $F$ is applied to the extremity of the cantilever (steel slide of size $1.9$cm $\times 1.4$cm $\times 100\mu$m) using a CSM Instruments Micro Indentation Tester (MHT); the deflection $\delta$ is measured using a camera. (b) (Left) Sketch of the setup where a drying drop, sitting on a PET slide, is approached to the cantilever. (Right) Superposition of images corresponding to zero and highest deflection
needed for drop slippage.}
\label{fig:friction}
\end{figure}

\begin{multicols}{2}
From these results, one can estimate the critical stress at cracking provided that the friction stress is known. Frictional responses of a system is a complex problem as it can be considered from different length scales, from macroscopic to molecular contact. Thus, the friction depends in a complicated way of the history of the system, the chemical-physics and the surface properties of the substrate. 
For all that reasons, measurements of friction force require a high level of expertise \cite{Gong2006}. 
In addition, in the present case,  the system evolves with time as consolidation proceeds. 
Thus we recognize that we can provide only an estimation of the gel friction force.

To do so, we use a cantilever technique calibrated by micro indentation testing as shown in figure \ref{fig:friction}-a.
Then, a colloidal drop is deposited with a syringe on a PET slide with an initial contact angle of $64^\circ$ and an initial volume of $2.7 mm^{3}$.
During the drying, a solid foot builds up and the center of the drop is a gel-like phase.
After $25$ minutes, a slight delamination of the drop foot occurs, preventing the pinning of the contact line while the center is still adhering on the substrate.
The PET slide is then translated quasistatically with a micrometric screw to push the sessile drop to the vertical cantilever figure (\ref{fig:friction}-b).
Thus the highest deflection $\delta$ resulting in the slippage of the drop is measured and the corresponding critical force is deduced. 
We assume then that this critical force is the sliding force. 
Our measurements are well reproducible allowing for a reliable estimation of the friction force $F_s$.
The experiments were repeated for solutions $\textrm{sol}_1$ and $\textrm{sol}_3$ giving respectively $F_s = 42\pm5$ mN and $F_s = 38\pm5$ mN.
Within the uncertainties, the friction force does not depend on glycerol concentration (for the low glycerol concentrations considered here) and is comparable to measurements of Vigil \textit{et al.} for similar systems of silica gels on hydrophobic surfaces \cite{Vigil1994}.

\end{multicols}
\begin{figure}
\centering
\resizebox{0.5\columnwidth}{!}{
  \includegraphics{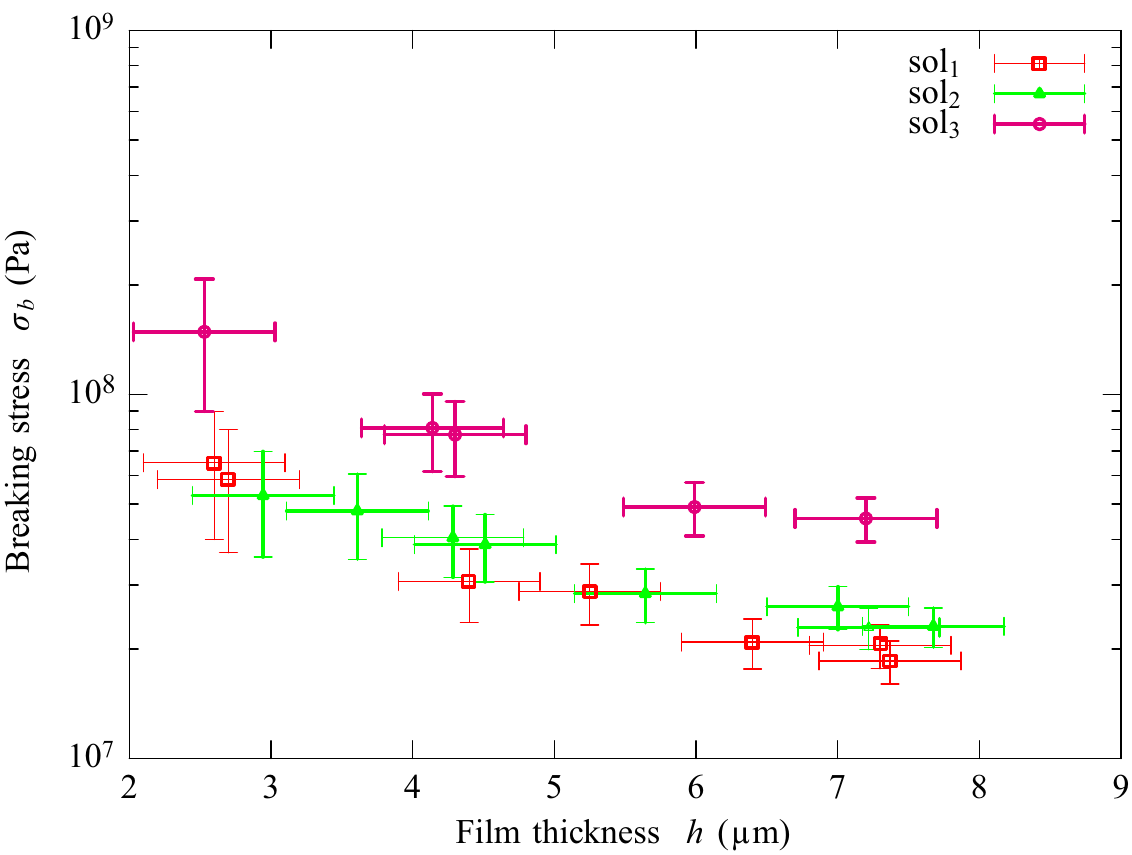}
}
\caption{Breaking stress $\sigma_b$  as a function of the final film thickness $h$ for the three colloidal dispersions.
The values are deduced from slopes in figure \ref{fig:wavelength} and by estimating the sliding stress $\sigma_s = F_s / (2 \pi R h)$ with $F_s=40$mN.
}\label{fig:stress}
\end{figure}

\begin{multicols}{2}

Taking $F_s=40$ mN, we can estimate the breaking stress $\sigma_b$ for each material with equation (\ref{eq:lambda_bar}) and the experimental data presented in figure \ref{fig:wavelength}.
The results are reported in figure \ref{fig:stress} showing values between typically $10$ to $100$ MPa.
At a given film thickness, the increase of the breaking stress with the glycerol concentration is coherent with a previous study focussed on the influence of a non-volatile compound added to a drying colloidal dispersions on the crack formations \cite{Boulogne2012a}.
This work revealed that due to additional mixing fluxes, the tensile stress responsible for cracks in the material is reduced and leads to an increase of the crack spacing as shown in figure \ref{fig:wavelength}.
Consequently, the breaking stress which stands for the macroscopic stress applied to the layer to break the material is higher for higher glycerol concentrations.

Moreover, for a given system, $\sigma_b$ decreases as the film thickness increases.
Man and Russel have reported experimental results in accordance with the decrease of $\sigma_b$ in respect of $h$ for thick films ($0.1$ to $1$mm) of silica and polystyrene particles \cite{Man2008}.
By extrapolating their data for silica particles to the thickness range of our films, cracking stress is found to be of the order of $10$ MPa which is close to our predictions.

To conclude, we have characterized the annular crack pattern of thin films of colloidal dispersions drying on a fiber.
This open geometry is of particular interest as it displays no boundaries and can be considered as one dimensional.
The spacing of annular cracks evolves linearly with the film thickness and depends on the mechanical properties of the gel.
We have interpreted our data with the Kitsunezaki model taking into account the general possibility for the gel to slip on the substrate.
The linear dependence of crack spacing on film thickness is found to be equal to the ratio between the critical cracking stress and the critical slipping stress.
From estimations of the friction force of colloidal gels on a hydrophobic substrate, we deduced the critical macroscopic stress required to break the materials.
Further experiments are now required concerning the direct measurements for the sliding force of these materials with hydrophobic surfaces.
\paragraph{acknowledgments}
We thank A. Aubertin, P. Jenfer, G. Marteau, and R. Pidoux for engineering and technical support and B. Cabane, U. Laujay and E. Mittelstaedt for valuable discussions.
We also thank Triangle de la Physique for the Anton Paar MCR 501 rheometer.

\end{multicols}
\end{document}